\documentclass[twocolumn]{aastex701}

\usepackage{mathtools}
\usepackage[capitalise]{cleveref}
\usepackage{threeparttable}
\usepackage{multirow}
\usepackage{booktabs}
\usepackage{amsmath}
\usepackage{graphicx}
\usepackage{natbib}
\usepackage{caption}

\begin{document}

\shorttitle{Evolution of Cosmic Voids}
\shortauthors{S. Tavasoli}

\title{Evolution of Cosmic Voids: Structure, Galaxies, and Dynamics}

\author[orcid=0000-0003-0126-8554,sname='Saeed Tavasoli']{Saeed Tavasoli}

\affiliation{Department of Astronomy and High Energy Physics, Kharazmi University, Tehran, Iran}
\email[show]{stavasoli@khu.ac.ir}

\begin{abstract}
We investigate the structural, photometric, and dynamical evolution of cosmic voids and their galaxy populations from $z=2.09$ to the present, focusing on void size as a key evolutionary parameter. Using void catalogs from four Millennium Simulation snapshots and SDSS data at $z<0.04$, we perform a unified analysis of void demographics, galaxy properties, and internal kinematics. Our analysis reveals clear evidence that cosmic voids exhibit a significant evolutionary trend of becoming progressively emptier toward low redshift, accompanied by a marked decline in the brightness and clustering of their galaxy populations.The void galaxy luminosity function evolves significantly: $M^{*}$ fades and $\alpha$ flattens with time; large voids host brighter galaxies, while small voids show stronger evolutionary changes. Stacked density profiles exhibit a universal shape when scaled by void radius, deepening and building more pronounced walls toward $z=0$. Galaxy spatial distributions reveal persistent size-dependent segregation, with galaxies in large voids lying farther from the center and more strongly clustered. Dynamical analysis of simulations shows coherent outward flows in all voids, with amplitudes decreasing toward $z=0$, providing a physical basis for observed redshift-space distortions. Comparison with SDSS broadly confirms these evolutionary trends but uncovers a non-zero central galaxy population in observed voids---absent in $\Lambda$CDM predictions---that may challenge current galaxy formation models in extreme underdensities. Future comparisons with additional simulations and deeper high-redshift surveys will provide stronger tests of $\Lambda$CDM in the most underdense regions.
\end{abstract}


\section{Introduction} \label{sec:intro}

The large-scale structure of the Universe manifests as a complex, interconnected web of galaxy clusters, filaments, sheets, and vast underdense regions known as cosmic voids \citep{gregory1978, bond1996}. This pervasive network is a natural consequence of the \(\Lambda\)CDM paradigm, where primordial density fluctuations are amplified by gravitational instability. The anisotropic collapse of matter along the principal axes of the large-scale tidal field sculpts the mass distribution into this web-like architecture \citep{zeldovich1970, shandarin1989}. Over cosmic time, this process gives rise to a hierarchy of environments with distinct dynamical properties, ranging from dense, virialized clusters at filament intersections to the underdense voids that dominate the Universe's volume.

Cosmic voids, first identified in early galaxy redshift surveys \citep{kirshner1981}, are expansive regions, typically tens of megaparsecs in radius, that occupy the majority of the cosmic volume while containing only a small fraction of its galaxies and matter \citep{dacosta1988, pan2012}. Governed primarily by large-scale expansion rather than non-linear collapse, their evolution is exquisitely sensitive to the underlying cosmology.

This makes voids potential cosmological probes, with studies suggesting their utility in constraining dark energy, modified gravity, and the mass of neutrinos \citep{hamaus2016, cautun2018};however, additional empirical validation is needed, particularly to verify how void density profiles, internal flows, and galaxy distributions map onto cosmological models. A diverse suite of void-based observables—including void abundances, density profiles, redshift-space distortions, imprints on the cosmic microwave background, Sunyaev--Zel'dovich signatures, weak lensing, baryon acoustic oscillations, the integrated Sachs--Wolfe effect, and the Alcock--Paczynski test—provides complementary constraints on both the expansion history and the
growth rate of structure \citep{granett2008, lavaux2012, ilic2013, sanchez2017, alonso2018, habibi2018, aubert2022, hamaus2022, kovacs2022, li2024}.

The identification of cosmic voids in both observational surveys and numerical simulations relies on a variety of algorithms, each with distinct definitions of what constitutes a void and different approaches to boundary delineation. Broadly, these methods fall into several categories: (i) void finders based on the watershed transform, which partition the density field into basins separated by ridges and naturally capture the hierarchical structure of the cosmic web \citep{platen2007, neyrinck2008}; (ii) sphere-based methods that search for maximal, non-overlapping empty spheres in the galaxy distribution, such as the popular Void Finder algorithm \citep{Hoyle2002,Banerjee2016,Sparkling2026} and its variants \citep{sutter2015}; (iii) dynamic approaches that trace the evolution of underdense regions in simulations using particle trajectories \citep{Colberg2005,Lavaux2009,Sartori2026}; and (iv) algorithms that identify voids directly from the Delaunay tessellation of the galaxy field \citep{schaap2000}. More recently, a novel approach based on genetic algorithms--VEGA (Voids idEntification using Genetic Algorithm)--has been introduced, which combines Voronoi tessellation with luminosity density contrast to identify void candidates \citep{ghafour2025}. While the specific void catalogs produced by different methods can vary in their detailed properties, the statistical trends--such as the void size function, density profiles, and evolutionary patterns--have been shown to be remarkably robust across algorithms, provided consistent selection criteria are applied \citep{Colberg2008, cautun2018}. In this work, we employ the void catalog generated by the algorithm of \citet{Aikio1998} on both the Millennium Simulation snapshots and the SDSS data; a detailed description of the void identification procedure and the construction of the catalogs is provided in Section~\ref{sec:voidcatalog}.

Concurrently, the underdense interiors of voids offer a unique laboratory for studying galaxy formation and evolution in relative isolation, largely shielded from the complex environmental effects prevalent in denser regions like clusters and filaments \citep{vandeweygaert2011,rodriguezmedrano2025}. Observations have consistently shown that void galaxies exhibit distinct properties compared to their counterparts in average or dense environments. They tend to have lower stellar masses \citep{hoyle2005, moorman2015}, bluer colors and later morphological types \citep{rojas2004, patiri2006, Hoyle2012, Tavasoli2015}, and are more gas-rich with lower gas-phase metallicities \citep{kreckel2015, douglass2017b}. This observed population, however, presents a challenge to \(\Lambda\)CDM predictions, which have historically struggled to reproduce the abundance of low-mass galaxies in these extreme underdensities—a discrepancy termed the ``void phenomenon'' \citep{Peebles2001, tikhonov2009, Tavasoli2013}.

Despite significant progress, there is still a lack of comprehensive understanding regarding the specific mechanisms and data that illustrate how the physical properties of voids and their galaxy populations co-evolve across cosmic time, particularly in terms of luminosity function variations and density profiles. In particular, the role of void size as a fundamental parameter governing both internal void dynamics and the evolutionary paths of their resident galaxies is not yet fully characterized. This paper aims to address this gap by performing a unified analysis of the structural, photometric, and dynamical evolution of voids and their galaxies, from high redshift to the present day. Using the high-resolution Millennium Simulation and comparing its predictions with observational data from the Sloan Digital Sky Survey (SDSS), we establish a coherent framework in which void size, cosmic epoch, and local environment jointly regulate the evolution of these underdense regions. 

This paper is organized as follows. In Section~2, we describe the construction of our
void--galaxy samples, including the Millennium Simulation dataset, the SDSS observational catalog,
and the void--finding procedure used to identify Small and Large voids. Section~3 presents our main
results, beginning with the luminosity function of void galaxies, followed by their radial
number--density profiles, the evolution of their spatial distributions, and finally the radial
velocity profiles that trace the dynamical state of voids across cosmic time. Finally, Section~4
summarizes our conclusions and discusses the broader implications of our findings for void evolution
and for tests of the $\Lambda$CDM cosmological model.

\begin{figure*}[ht!]
    \centering
    \includegraphics[width=\textwidth]{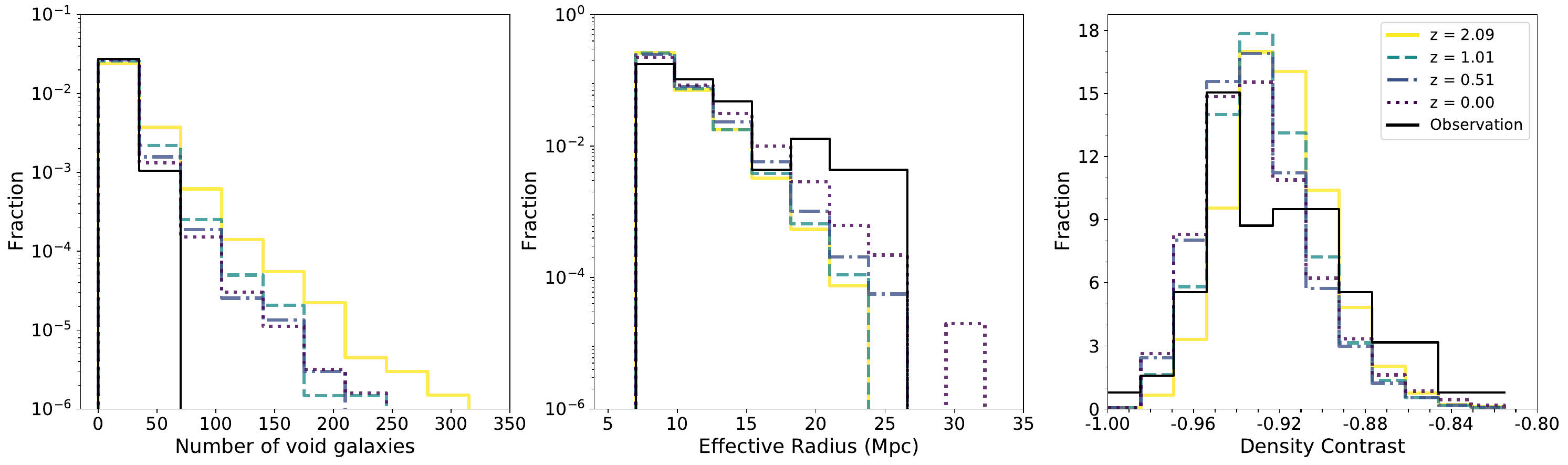} 
    \caption{Statistical overview of our void catalogs: distributions of void galaxy counts, effective radii, and density contrasts at four simulation redshifts (\(z=0.0,\,0.51,\,1.01,\,2.09\)) together with the observational sample at \(z\sim0\).}
    \label{fig:hist}
\end{figure*}

\begin{figure}[ht!]
    \centering 
        \includegraphics[width=\columnwidth]{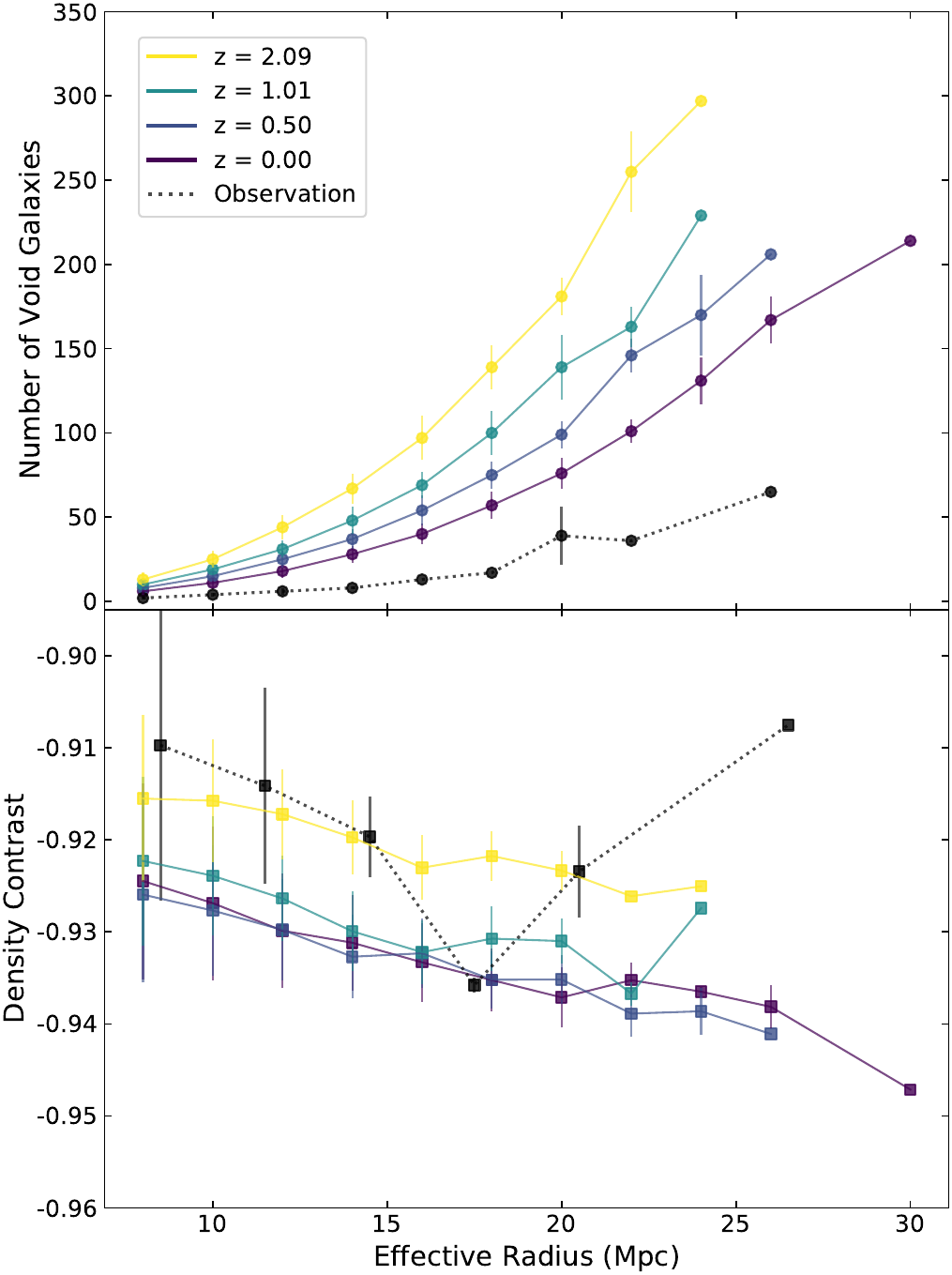}
        \caption{Top: mean number of galaxies per void as a function of void effective radius \(R_{v}\). Bottom: mean density contrast \(\langle\delta_{v}\rangle\) versus \(R_{v}\). Points denote bin-averaged values and vertical bars indicate the standard error of the mean ($\pm 1\sigma$). \textit{Note that \(\delta_{v}\) is computed relative to the sample background density ($\rho_{b}\sim 0.011$) for SDSS and ($\rho_{b}\sim 0.039$) for the simulation); this difference contributes to the apparent offset between the datasets.}}

        \label{fig:scatter}  
\end{figure}

\section{Sample Selection} \label{sec:sample}

To investigate the evolutionary behavior of galaxies residing in under-dense environments, we employ both simulated and observational datasets. The simulated sample is drawn from the Millennium cosmological simulation \citep{Springel2005}, which provides a large-volume, high-resolution framework suitable for tracing the statistical properties of large-scale structure across cosmic time. To complement the simulation and to anchor our results at the present epoch, we additionally incorporate an observational void galaxy catalog constructed from the Sloan Digital Sky Survey (SDSS). The details of both datasets are described below.

\subsection{Millennium Simulation Sample} \label{sec:Millennium}

To trace the evolution of galaxies in cosmic voids, we use the semi-analytic galaxy catalog of \citet{Guo2011}, implemented on the Millennium Run simulation \citep{Boylan-Kolchin2009}. The simulation adopts a flat $\Lambda$CDM cosmology consistent with the WMAP7 results \citep{Komatsu2011}, with parameters:
$\Omega_{\Lambda} = 0.728$,
$\Omega_{m} = 0.272$,
$\Omega_{b} = 0.045$,
$h = 0.70$,
$n = 0.96$,
and $\sigma_{8} = 0.807$.
The computational domain is a periodic cube of $500 \, h^{-1} \, \mathrm{Mpc}$ on each side, resolved with $2160^{3}$ particles. We analyze four simulation snapshots at redshifts $z = 0.0$, $0.51$, $1.01$, and $2.09$, providing a well-spaced sampling from the intermediate-redshift universe to the present day. This enables us to follow the secular evolution of void galaxies over several billion years. To ensure completeness across all epochs, we impose a stellar mass threshold of $9.31 \times 10^{8}\, M_{\odot}$, consistent with the resolution limit of the simulation. Additionally, we select all galaxies brighter than approximately $M_{r} \sim -18$ in the $r$-band, allowing a direct comparison with the observational dataset described below. After applying these criteria, the resulting samples contain $\sim 5 \times 10^{6}$ , $\sim 7 \times 10^{6}$, $\sim 9 \times 10^{6}$, and $\sim 11 \times 10^{6}$ galaxies at $z=0.0$, $z=0.51$, $z=1.01$, and $z=2.09$, respectively.

\subsection{Observational Sample} \label{sec:SDSS}

To characterize the distribution of galaxies in observed under-dense environments, we adopt the complete void catalog of \citet{Tavasoli2021} from the spectroscopic sample of SDSS DR16 \citep{Ahumada2020} (A detailed description follows in the next section). The selected region of the sky spans $130^\circ < \mathrm{R.A.} < 235^\circ$ and $0^\circ < \mathrm{decl.} < 55^\circ$, containing approximately 40,000 galaxies up to $z \sim 0.04$. All galaxy redshifts are corrected to the cosmic microwave background (CMB) rest frame to ensure accurate distance estimates. K-corrections are applied using the algorithm developed by \citet{Blanton2003} and \citet{Blanton2007}. To construct a homogeneous, volume-limited sample suitable for statistical analysis, we adopt an absolute magnitude limit of $M_{r} \sim -18$, consistent with the selection applied to the simulated galaxies. This magnitude cut defines the upper redshift limit of $z \sim 0.04$ and yields a final sample of approximately 30,000 galaxies. The absolute magnitudes are computed assuming $h = 0.70$, $\Omega_{\Lambda} = 0.73$, and $\Omega_{m} = 0.27$, which are fully consistent with the cosmological parameters of the Millennium simulation. This consistency ensures that the comparison between the simulated and observed void galaxy populations at $z \sim 0$ is physically meaningful and free from systematic offsets arising from differing cosmological assumptions.

\subsection{Void Catalog Construction} \label{sec:voidcatalog}

In this study, we adopt the three-dimensional implementation of the \citet{Aikio1998} algorithm, as extended by  \citet{Tavasoli2013}, which has been extensively validated in previous void studies and does not impose any assumption of spherical symmetry on voids (see \citealt{Colberg2008} for a review of void-finding techniques). Applying this same procedure to both the Millennium simulation and the SDSS sample ensures methodological consistency and enables a direct comparison of their void populations. Prior to applying the algorithm on a Cartesian grid, galaxies are classified as wall or field galaxies based on their nearest-neighbor distances following \citet{Hoyle2002}. Field galaxies are considered candidates for void membership, while the Aikio-M{\"a}h{\"o}nen algorithm constructs a distance field on the gridded wall-galaxy distribution. Each grid element is then assigned to a subvoid using the climbing algorithm of \citet{Schmidt2001}. Two subvoids are merged into a larger void if the separation between them is smaller than their corresponding distance-field values. The void volume is computed from the number of grid cells associated with each void multiplied by the cell volume (see \citealt{Tavasoli2013} for further details). The void center is defined as the centroid of the grid points enclosing the elementary cell.

The resulting void catalogs span a wide range of void sizes, characterized by the effective radius $R_{v}$ and the number-density contrast $\delta_{v}$. The effective radius is defined as the radius of a sphere with the same volume as the void. The number-density contrast is defined as
\begin{equation}
\delta_{v} = \frac{\rho_{v} - \rho_{b}}{\rho_{b}},
\label{eq:density}
\end{equation}
where $\rho_{v}$ is the number density of galaxies inside the void and $\rho_{b}$ is the mean number density of the corresponding sample, ensuring that $\delta_{v}$ reflects the relative underdensity of the void compared to the background density. Hereafter, "density contrast" refers to the number-density contrast  $\delta_{v}$.

For the observational SDSS dataset, the same void-finding procedure is applied to the volume-limited galaxy catalog described in Section~\ref{sec:SDSS}. The SDSS footprint has an irregular geometry and limited sky coverage, and the combination of this restricted volume with the magnitude cut of $M_{r} < -18$ results in a substantially smaller number of galaxies and voids compared to the simulated samples. This reduced sampling naturally leads to larger statistical uncertainties in the derived void properties, a factor we account for when interpreting the observational results. To avoid spurious detections, we exclude voids smaller than $R_{v} < 7 \, h^{-1}\,\mathrm{Mpc}$, and to minimize boundary-related biases in the inferred galaxy positions, we retain only those voids that lie fully within the geometrical limits of each dataset either the simulation box or the SDSS survey footprint.

Our final void catalogs include four simulation snapshots at redshifts $z = 0.0$, $0.51$, $1.01$, and $2.09$, and one observational catalog at $z \sim 0$ from SDSS. Statistical properties of all catalogs including the number of voids ($N_{\mathrm{void}}$), the number of void galaxies ($N_{\mathrm{gal}}$), the mean effective radius ($\overline{R_{v}}$), and the mean density contrast ($\overline{\delta_{v}}$) are summarized in Table~\ref{tab:voidcatalog}. The substantially smaller size of the observational catalog, as reflected in Table~\ref{tab:voidcatalog}, should be kept in mind when comparing the corresponding $z \sim 0$ results.

\begin{table}[ht!]
\centering
\caption{Properties of the void catalogs for the Millennium simulation and the SDSS observational sample.}
\vspace{-0.5em}
\label{tab:voidcatalog}
\small
\setlength{\tabcolsep}{4pt} 
\begin{tabular}{ccccc}
\hline
Dataset & $N_{\mathrm{void}}$ & $N_{\mathrm{gal}}$ & $\overline{R_{v}}$ [Mpc] & $\overline{\delta_{v}}$ \\
\hline
\multicolumn{5}{c}{\text{Millennium Simulation}} \\
$z=0.00$ & 17873 & 223818 & $9.7 \pm 2.5$ & $-0.935 \pm 0.03$ \\
$z=0.51$ & 19068 & 274990 & $9.3 \pm 2.1$ & $-0.930 \pm 0.02$ \\
$z=1.01$ & 19432 & 332027 & $9.0 \pm 1.9$ & $-0.927 \pm 0.02$ \\
$z=2.09$ & 19091 & 435667 & $9.0 \pm 1.9$ & $-0.915 \pm 0.02$ \\
\hline
\multicolumn{5}{c}{\text{SDSS Observation}} \\
$z \sim 0$ & 93 & 512 & $10.35 \pm 3.3$ & $-0.926 \pm 0.04$ \\
\hline
\end{tabular}
\end{table}

\begin{figure*}[!htbp] 
\centering 
\includegraphics[width=\linewidth]{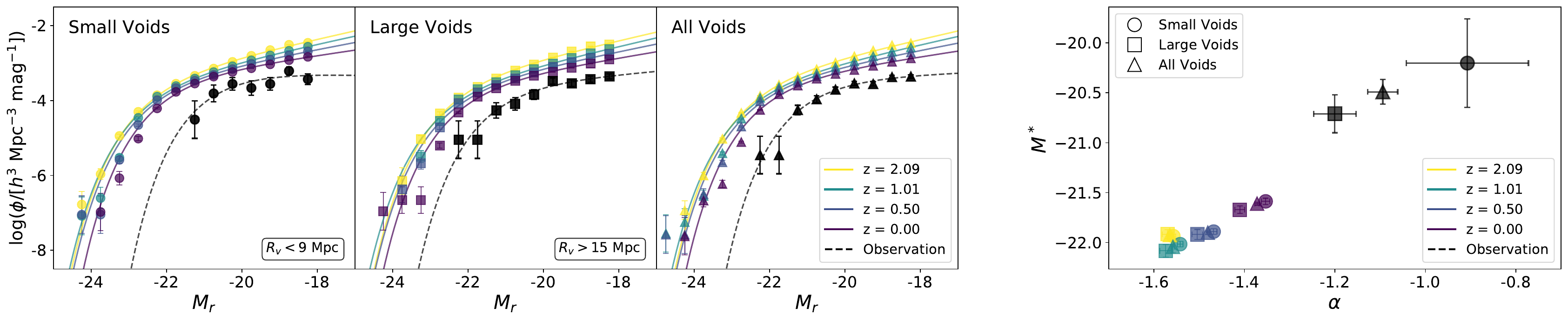}
    \caption{\textit{Left:} Galaxy luminosity functions for void galaxies in the \textit{Small}, \textit{Large}, and \textit{All} samples across four simulation redshift bins, with the SDSS $z\!\sim\!0$ measurements overplotted for comparison. \textit{Right:} Redshift evolution of the best-fit Schechter parameters $M^*$ and $\alpha$ for the simulated samples, highlighting systematic trends with void size and the offset between simulations and observations at $z\!\sim\!0$.}
    \label{fig:lf}
\end{figure*}

Figure~\ref{fig:hist} presents a statistical comparison of our void catalogs by showing the distributions of the void galaxy counts, effective radii, and density contrasts for the four simulation snapshots at $z = 0.0$, $0.51$, $1.01$, and $2.09$, together with the observational sample at $z \sim 0$. As seen in the simulation results, high-redshift voids contain a larger number of galaxies and exhibit higher number densities than their present-day counterparts, and the largest voids are absent at early cosmic times. These trends are consistent with the findings of \citet{Schuster2023} based on the Magneticum simulations and can be interpreted through the ``void-in-void'' hierarchical growth process \citep{Sheth2004}.

To quantify the evolutionary changes visible in the histograms, we perform a two-sample Kolmogorov--Smirnov test between the $z=0$ and $z=2.09$ snapshots for each of the three void properties. Given the large and comparable sizes of the two samples (see Table~\ref{tab:voidcatalog}), the test confirms that the distributions of void galaxy counts, effective radii, and density contrasts are all significantly different ($p<0.001$ for each property), supporting the evolutionary trends discussed above.

The observational distributions provide an additional point of comparison at $z \sim 0$. The number of void galaxies in the SDSS sample is substantially lower than in the simulated $z=0$ snapshot, reflecting the well-known ``void phenomenon'' highlighted by \citet{Peebles2001}, in which the observed abundance of galaxies inside voids is smaller than predicted by the standard $\Lambda$CDM model. The distribution of effective radii shows broad statistical agreement between the observational and simulated samples; however, the simulated histogram exhibits a modest excess at the largest radii, corresponding to roughly one histogram bin width. Given the limited size of the observational catalog and the associated sampling uncertainties, this small offset does not imply a clear tension. Finally, similar to the simulations, the observational voids at $z \sim 0$ tend toward more negative density contrasts, indicating that voids in both datasets become increasingly underdense at the present epoch, with an overall statistical correspondence between the two catalogs.

Figure~\ref{fig:scatter} shows the relation between the void effective radius \(R_{v}\) and (i) the mean number of void galaxies \(\langle N_{\rm gal}\rangle\) and (ii) the mean density contrast \(\langle\delta_{v}\rangle\). For each radius bin we plot the bin-averaged values with vertical error bars indicating the standard error of the mean (SEM). The simulated snapshots exhibit a clear trend in which the number of galaxies per void increases toward higher redshift across the full radial range. Furthermore, at a given radius the density contrast of simulated voids is greater (less negative) at high redshift compared to low redshift.

The observed monotonic decline in void galaxy population from high to low redshift requires a physical explanation. A plausible mechanism is the gradual evacuation of galaxies from void interiors due to their gravitational attraction toward, and subsequent merging with, surrounding higher-density structures such as filaments and sheets. In-situ galaxy--galaxy mergers within voids may also reduce number counts over time; however, given the intrinsically low galaxy density in void environments, the efficiency of in-situ mergers is likely subdominant relative to the large-scale dynamics of the cosmic web.

When the SDSS sample is included for comparison at \(z\!\sim\!0\), the observational catalog shows systematically lower galaxy counts than the simulated \(z=0\) snapshot in every radius bin. In the \(\delta_{v}\)–\(R_{v}\) panel the observational voids display systematically higher (less negative) density contrasts than the simulated voids across the entire radius range, i.e., the simulated voids are on average more underdense. This apparent offset is primarily driven by differences in the effective background densities used to define the density contrast: the SDSS volume‑limited sample has $\rho_{b}\sim 0.011$, whereas the simulated snapshot has $\rho_{b}\sim 0.039$. Owing to the fact that \(\delta_{v}\)  is normalized to the sample-specific mean background density (see Equation~\eqref{eq:density}), the resulting density contrasts are intrinsically tied to the selection function of each catalog and thus are not directly comparable, even when a uniform magnitude limit $M_{r} < -18$ is adopted.

We therefore retain the \(\delta_{v}\)–\(R_{v}\) plot for completeness but caution that the observed offset primarily reflects background normalization differences rather than an unambiguous physical inconsistency between the observational and simulated void populations; the relation should be interpreted conservatively. Reporting bin-averaged values and SEMs makes the comparison transparent and allows the reader to assess the statistical significance of any residual offsets after accounting for background differences.

\subsection{Systematic Differences Between Observational and Simulated Void Catalogs} \label{sec:Systematicdifferences}
Although the same absolute--magnitude limit and void--finding algorithm are applied to both the Millennium simulation and the SDSS dataset, several intrinsic differences between observational and simulated samples remain and may influence the comparison of void and galaxy properties. These systematics arise from the fundamentally distinct nature of real and synthetic galaxy distributions and must be considered when interpreting the results.
First, the SDSS catalog is affected by redshift--space distortions (RSDs), which elongate structures along the line of sight and can modify the inferred shapes and sizes of voids. The simulated catalogs, in contrast, provide a complete real--space distribution of galaxies unless RSDs are explicitly imposed. This difference can lead to systematic offsets in the effective radii and density contrasts of voids identified in the two datasets.
Second, the survey geometry and limited sky coverage of SDSS introduce boundary effects that are absent in the periodic simulation box. Although we explicitly remove all voids intersecting the survey boundaries in both datasets--thereby eliminating the most direct form of edge contamination--the effective survey volume of SDSS remains substantially smaller and geometrically irregular. This reduced volume enhances cosmic variance, suppresses the detection of the largest voids, and alters the merging of subvoids in ways that cannot be fully mitigated by boundary cuts alone.
Third, the SDSS sample is subject to spectroscopic incompleteness and fiber--collision losses, which primarily affect high--density environments such as galaxy groups and clusters. Since our void--finding algorithm explicitly removes overdense regions and focuses on isolated galaxies within underdense environments, the impact of fiber collisions on the present analysis is expected to be minimal. The density field relevant for void identification is therefore only weakly influenced by these observational limitations.
Fourth, the lower galaxy number density of SDSS introduces shot noise, which can in principle suppress the detection of small voids. However, the void--radius distribution presented in Figure~\ref{fig:hist} shows that the abundance of small voids is comparable between the observational and simulated catalogs, while the dominant differences arise at the large--radius end. This indicates that shot noise does not significantly bias the identification of small voids in our SDSS sample and primarily affects the statistics of the largest voids.
Finally, the semi--analytic model used to populate the Millennium simulation introduces its own astrophysical systematics, including uncertainties in galaxy luminosities, star--formation histories, and satellite fractions. These modeling uncertainties may affect the abundance and internal properties of void galaxies in ways that differ from the observational sample, even when identical magnitude cuts are applied.
Taken together, these observational and modeling systematics do not compromise the qualitative comparison between the simulated and observed void populations, but they do introduce quantitative differences that must be kept in mind when interpreting the relative amplitudes of void radii, density contrasts, and galaxy counts. These additions provide a clearer and more comprehensive explanation of the methodological differences between the two datasets and strengthen the interpretation of the results presented in the subsequent sections.

\section{ANALYSIS AND results} \label{sec:result}

Since the correlation of void properties (Sec.~\ref{sec:voidcatalog}) alone does not fully determine their kinematic content, this section presents a detailed analysis of the evolutionary trends for galaxies residing in cosmic voids across our four simulation snapshots. These trends can only be assessed for the simulated catalogs, where voids are available at multiple redshifts (\(z = 0.0,\,0.51,\,1.01,\,2.09\)). In addition to tracing the redshift evolution, we also compare the simulation results with the \(z\sim 0\) observational SDSS void sample to evaluate the degree of consistency between simulated and real void environments.

To investigate the potential dependence of galaxy properties on void size, we classify voids in each simulation snapshot into three categories based on their effective radius \(R_{v}\): \textit{small voids} (\(R_{v}<9~\mathrm{Mpc}\)), \textit{large voids} (\(R_{v}>15~\mathrm{Mpc}\)), and the full \textit{all-void} population, which serves as a reference. The same size-based classification is applied to the SDSS observational voids at \(z\sim 0\), enabling a consistent comparison between the simulated and observational samples. The resulting sample sizes for all categories in both datasets are summarized in Table~\ref{tab:voidsize}.

As reported in Table~\ref{tab:voidsize}, the simulated catalogs exhibit a clear evolutionary trend: toward lower redshift, the number of \textit{small} voids decreases while the number of \textit{large} voids increases. This behavior is qualitatively consistent with hierarchical void growth and merging, in which small voids coalesce or are absorbed into larger underdense regions as structure evolves (see \citealt{Sheth2004}). The SDSS observational sample at \(z\!\sim\!0\) provides an empirical anchor for the low-redshift population, although the observational data do not allow direct tracking of void evolution across time.

\begin{table}[!htbp]
\centering
\caption{Counts of voids ($N_{\rm void}$) and void galaxies ($N_{\rm galaxy}$) by void-size category and redshift. Small: $R_{v}<9\ \mathrm{Mpc}$; Large: $R_{v}>15\ \mathrm{Mpc}$; All: full void population.}
\label{tab:voidsize}
\scriptsize
\renewcommand{\arraystretch}{0.95}

\begin{tabular}{l
    @{\hskip18pt}  r@{\hskip 4.2pt}r   
    @{\hskip20pt} c@{\hskip 2.2pt}l   
    @{\hskip10pt} r@{\hskip 4.2pt}r}  

\toprule
Sample &
\multicolumn{2}{l}{Small Voids} &
\multicolumn{2}{l}{Large Voids} &
\multicolumn{2}{c}{All Voids} \\
\cmidrule(l{2pt}r{18pt}){2-3}
\cmidrule(l{2pt}r{12pt}){4-5}
\cmidrule(l{2pt}r{4pt}){6-7}
 & $N_{\rm void}$ & $N_{\rm galaxy}$
 & $N_{\rm void}$ & $N_{\rm galaxy}$
 & $N_{\rm void}$ & $N_{\rm galaxy}$ \\
\midrule

\multicolumn{7}{c}{Millennium simulation} \\
$z=0.00$ & 9065  & 55517  & 818 & 42903  & 17873 & 223818 \\
$z=0.51$ & 10770 & 85889  & 468 & 30853  & 19068 & 274990 \\
$z=1.01$ & 11717 & 119151 & 326 & 26310  & 19432 & 332027 \\
$z=2.09$ & 11644 & 157076 & 286 & 32830  & 19091 & 435667 \\

\midrule

\multicolumn{7}{c}{SDSS observation} \\
$z\!\sim\!0$ & 40 & 63 & 6 & 192 & 93 & 512 \\
\bottomrule

\end{tabular}
\end{table}

For each void-size category and redshift, we analyze and compare key galactic properties, including the galaxy luminosity function, the internal radial number-density profile, and the kinematic properties. This comparative approach enables us to trace how both cosmic epoch and host-void scale influence galaxy evolution within underdense environments.

\subsection{Luminosity Function}

The luminosity function (LF) is a key statistical tool for studying galaxy populations in underdense environments. Its evolution provides critical insight into how galaxy formation proceeds in cosmic voids, offering a test for theoretical models (e.g., \citealt{berlind2002}). The influence of void environments on the LF has been investigated using observational data, simulations, and theoretical frameworks (e.g., \citealt{grogin1999, hoyle2005, tempel2011, moorman2015, curtis2024, ashurikisomi2025}). For instance, \citet{Tavasoli2015} found that the LF in moderately underdense voids ($\delta > -0.87$) is well described by a Schechter function, whereas this form breaks down in the most rarefied voids ($\delta < -0.95$).

In this work, we measure the redshift evolution of the $r$-band LF for void galaxies from $z = 0$ to $z = 2.09$ using the galaxy catalogs from the Millennium simulation. For each of the four simulation snapshots (corresponding to $z = 0.0, 0.51, 1.01, 2.09$), we compute the luminosity function separately for three void categories: the complete population (All voids), as well as \textit{Small} voids ($R_v < 9\ \text{Mpc}$) and \textit{Large} voids ($R_v > 15\ \text{Mpc}$).

To complement the simulated LFs, we also compute the $r$-band luminosity function for the SDSS void galaxies at \(z\!\sim\!0\), using the same void-size classification applied to the simulation. Although the SDSS sample provides only a single-epoch measurement, it serves as an essential observational benchmark for assessing the realism of the simulated LF at low redshift. The SDSS LF additionally allows us to test whether the relative differences between Small and Large voids observed in the simulations are also present in real data, despite the observational limitations imposed by survey geometry and magnitude completeness.

The luminosity function, $\phi(M)$, constructed from our simulated and observational void samples, is parameterized by the Schechter function \citep{schechter1976}:

\begin{equation}
\begin{multlined}
\phi(M) dM = 0.4 \ln(10) \phi^* 10^{0.4(\alpha+1)(M^*-M)} \\
\times \exp \left( -10^{0.4(M^*-M)} \right) dM
\end{multlined}
\end{equation}

where $M^{*}$ is the characteristic luminosity, $\alpha$ is the faint-end slope, and $\phi^{*}$ is the normalization.

\begin{table*}[t!]
\centering
\caption{Best-fit Schechter parameters ($M^{*}$, $\alpha$) with $1\sigma$ uncertainties for galaxies in the Small, Large, and All void categories from the simulation snapshots and the SDSS \(z\!\sim\!0\) smaple .}
\label{tab:schechter_fits}
\renewcommand{\arraystretch}{0.95}

\begin{tabular*}{0.98\textwidth}{@{\extracolsep{\fill}} l
    @{\hskip6pt} c@{\hskip2pt}c    
    @{\hskip18pt} c@{\hskip2pt}c   
    @{\hskip18pt} c@{\hskip2pt}c } 
\toprule
Sample &
\multicolumn{2}{c}{Small Voids} &
\multicolumn{2}{c}{Large Voids} &
\multicolumn{2}{c}{All Voids} \\
\cmidrule(l{4pt}r{18pt}){2-3}\cmidrule(l{4pt}r{18pt}){4-5}\cmidrule(l{4pt}r{4pt}){6-7}
 & $M^{*}$ & $\alpha$ & $M^{*}$ & $\alpha$ & $M^{*}$ & $\alpha$ \\
\midrule

\multicolumn{7}{c}{Millennium simulation} \\
$z=0.0$  & $-21.58\pm0.03$ & $-1.35\pm0.01$   & $-21.67\pm0.03$ & $-1.41\pm0.01$  & $-21.61\pm0.01$ & $-1.37\pm0.01$   \\
$z=0.51$ & $-21.89\pm0.03$ & $-1.46\pm0.01$   & $-21.91\pm0.04$ & $-1.50\pm0.01$  & $-21.90\pm0.01$ & $-1.48\pm0.01$   \\
$z=1.01$ & $-22.01\pm0.02$ & $-1.54\pm0.01$   & $-22.08\pm0.06$ & $-1.57\pm0.01$  & $-22.04\pm0.06$ & $-1.55\pm0.01$   \\
$z=2.09$ & $-21.93\pm0.02$ & $-1.55\pm0.01$   & $-21.91\pm0.04$ & $-1.57\pm0.01$  & $-21.92\pm0.01$ & $-1.56\pm0.01$   \\

\midrule

\multicolumn{7}{c}{SDSS observation} \\
$z\!\sim\!0$ & $-20.20\pm0.4$ & $-0.91\pm0.15$   & $-20.71\pm0.18$ & $-1.20\pm0.05$  & $-20.49\pm0.11$ & $-1.09\pm0.03$  \\
\bottomrule
\end{tabular*}
\end{table*}

To overcome the low galaxy counts in individual voids and to obtain robust statistics, we combine all galaxies within each void-size category for a given snapshot. For both the simulated snapshots and the SDSS sample at $z\!\sim\!0$, the luminosity function is computed by dividing the number of galaxies in each magnitude bin by the total volume of the voids belonging to that category. Because our void catalogs include only fully enclosed voids, the effective void volumes are well defined in both datasets, enabling a consistent LF estimation across simulations and observations. This stacking procedure allows us to trace the LF over the absolute-magnitude range $-25 < M_{r} < -18$ and to measure the evolution of the faint-end slope $\alpha$ and the characteristic magnitude $M^{\ast}$ up to $z = 2.09$ for each void-size class, while simultaneously providing a direct observational benchmark at low redshift.

\begin{table}[htbp]
\centering
\setlength{\tabcolsep}{2pt} 
\small                      
\caption{Best-fit parameters $q$ and $r$ describing the evolution of $M^*$ and $\alpha$ with redshift
(see Equations~\ref{eq:mstar_evolution} and \ref{eq:alpha_evolution}) for each void category.}
\vspace{-1.0em}
\begin{tabular}{@{}l c c c c c@{}}
\hline
Void & $M^*(z_0=0)$ & $q$ & $\alpha(z_0=0.51)$ & $r$ \\
\hline
Small & -21.58 & $0.164\!\pm\!0.005$ & -1.46 & $0.048\!\pm\!0.02$ \\
Large & -21.67 & $0.121\!\pm\!0.001$ & -1.50 & $0.038\!\pm\!0.01$ \\
All   & -21.61 & $0.131\!\pm\!0.003$ & -1.48 & $0.042\!\pm\!0.01$ \\
\hline
\end{tabular}
\label{tab:evolution_fits}
\end{table}


The left panel of Figure~\ref{fig:lf} presents the $r$-band luminosity function for the three void
categories (Small, Large, and All voids) across multiple redshifts. Solid lines denote the best-fit
Schechter functions obtained via least-squares fitting to the stacked void samples, while the SDSS
measurement at $z\!\sim\!0$ is overplotted as a black dashed line to provide a direct low-redshift
benchmark. The characteristic Schechter form—a power-law faint end dominated by low-mass galaxies
and an exponential cutoff at the bright end populated by rare, massive systems—is evident in both the
simulations and the SDSS data. As expected for underdense environments, bright galaxies are scarce
and the population is dominated by fainter objects \citep{rojas2004}. The corresponding best-fit
Schechter parameters and their $1\sigma$ uncertainties are listed in Table~\ref{tab:schechter_fits}.
The overall similarity in shape across void categories indicates that the fundamental form of the
luminosity function is preserved in underdense regions, although the fitted parameters (notably the
faint-end slope $\alpha$ and characteristic magnitude $M^{\ast}$) exhibit systematic variations with void
size and redshift; these variations are discussed below in the context of the SDSS comparison.

The right panel of Figure~\ref{fig:lf} shows the correlation between the Schechter parameters $M^{*}$ and $\alpha$ for 
the void galaxy samples, revealing a clear evolutionary trend across the four simulation snapshots.
 For the All void population, the characteristic magnitude brightens from $M^{*} = -21.61 \pm 0.01$ at
 $z = 0.0$ to $M^{*} = -21.92 \pm 0.01$ at $z = 2.09$, while the faint-end slope steepens
 from $\alpha = -1.37 \pm 0.01$ to $\alpha = -1.56 \pm 0.01$ over the same redshift range. 
 The Small-void and Large-void subsamples exhibit the same qualitative behavior. 
 This evolution toward a shallower faint-end slope (less negative $\alpha$) at lower redshifts
 indicates that the relative abundance of faint galaxies in voids decreases over cosmic time,
 consistent with the suppression of star formation in low-mass void galaxies at late epochs.

While this overall trend is coherent, quantifying its functional form reveals significant differences
between void sizes. To characterize these differences, we fit the redshift dependence of $M^*$ and
$\alpha$ using the empirical forms suggested by \citet{Helgason2012}:

\begin{align}
    M^*(z) &= M^*(z_0) - 2.5 \log \left[ (1 + (z - z_0))^q \right], \label{eq:mstar_evolution}\\
    \alpha(z) &= \alpha(z_0) \cdot \left( \frac{z}{z_0} \right)^r. \label{eq:alpha_evolution}
\end{align}

where $q$ and $r$ are the evolutionary indices to be fitted. Here, we set the pivot redshift $z_0 = 0$ for $M^*(z)$ and $z_0 = 0.51$ for $\alpha(z)$. The latter choice avoids the singularity at $z=0$ in Equation~(\ref{eq:alpha_evolution}) and places the pivot near the center of our redshift range, following the approach of \citet{Helgason2012}. The distinct best-fit parameters for each category (Table~\ref{tab:evolution_fits}) confirm that the characteristic luminosity fades and the faint-end slope flattens at markedly different rates, indicating that the host void's size modulates the pace of evolution.

In addition to the simulated trends, the SDSS measurement at $z\!\sim\!0$ provides an important
observational benchmark. Similar to the simulations, the observational data exhibit a clear dependence
of the Schechter parameters on void size: both $M^{*}$ and $\alpha$ are more negative in Large voids
than in Small ones, indicating that the void-size hierarchy observed in the simulations is also present
in real data. However, when compared directly to the simulated $z=0$ snapshot, the SDSS values are
systematically shifted toward less negative $M^{*}$ and $\alpha$ across all void categories. This implies
that, at fixed void size, the simulated void galaxy population is characterized by a brighter
characteristic magnitude and a steeper faint-end slope than observed. While part of this difference may
reflect the distinct background densities and selection functions of the two datasets, the systematic
nature of the offset suggests that the observational luminosity function provides a non-trivial
constraint on the low-redshift normalization of the simulated evolutionary trends. We note that the luminosity-function parameters are derived from least-squares fits to binned galaxy counts, a procedure that is sensitive to binning choices and does not capture the full covariance between $M^{*}$, $\alpha$, and $\phi^{*}$. Likelihood-based estimators such as the STY method \citep{Sandage1979} provide a more rigorous statistical treatment, but applying them lies beyond the scope of the present analysis. Notably, this behavior is consistent with the findings of \citet{Tavasoli2013}, who showed that for a given void radius the total luminosity of simulated void galaxies exceeds that of their observational counterparts.

This scale-dependent evolution aligns with a broad scenario of “void downsizing,” in which massive
galaxy formation ceases early in underdense regions while low-mass galaxies undergo passive fading
or depletion (e.g., \citealt{vanZee2006, moorman2015}). Our results extend this picture by
demonstrating that the efficiency of this downsizing process is itself environmentally dependent,
proceeding at different rates in voids of different sizes. Direct physical evidence for this downsizing
comes from the star formation properties of galaxies near the characteristic luminosity
($M^*_r \pm 0.3$ mag). For this sample, the average $g-r$ color reddens from $0.26 \pm 0.09$ at
$z \approx 2.09$ to $0.54 \pm 0.02$ at $z = 0.0$, while $\log(\mathrm{sSFR\,[yr^{-1}]})$ declines from
$-8.96 \pm 0.29$ to $-10.18 \pm 0.62$. This progressive quenching and reddening, observed across all
void categories, confirms that the stellar populations are aging and star formation is slowing down.
These findings from the Millennium simulation are in excellent agreement with observational
conclusions that void galaxies have slower star formation histories \citep{dominguezgomez2023}.
Together, the coordinated trends in $M^*_r$, $\alpha$, color, and sSFR paint a consistent picture of
void galaxies experiencing a gradual slowdown in star formation after an early epoch of more efficient
growth.

\subsection{Radial Number Density Profiles}

The large number of voids in our catalog enables a precise measurement of the average (stacked) radial number density profile. We investigate how these profiles evolve across cosmic time and how they differ between void size categories (Small, Large, and the complete All-voids sample). In addition to the simulated profiles, we also include the SDSS void galaxy sample at $z\!\sim\!0$ to provide an observational benchmark for assessing the low-redshift normalization of the simulated trends.

In the literature, various void-finding algorithms and methodologies for constructing average profiles have led to a diversity of proposed functional forms for the void density profile. However, a key finding from both observational and simulation studies is that when distances from the void center are scaled in units of the void radius ($r/R_v$), the average re-scaled density profile appears to be approximately universal—largely independent of void size and cosmic epoch (e.g., \citealt{ceccarelli2013, hamaus2014, nadathur2014, Tavasoli2021, schuster2025, Curtis2025Density}). The inclusion of the SDSS void sample allows us to test this universality directly at low redshift using real galaxies.

To construct our profiles, we must choose a density estimator. \citet{nadathur2014} evaluated three main estimators for stacked void profiles: the naive galaxy count, the Voronoi Tessellation Field Estimator (VTFE), and the Poisson estimator. The naive method can be systematically biased low in the centers of small voids, while VTFE relies on constructing a Voronoi tessellation, as used in the ZOBOV algorithm of \citet{neyrinck2008}. In this work, we adopt the Poisson estimator, which is well suited for our analysis and consistent with the underlying assumptions of the void-finding algorithm applied in this study (see \citealt{Tavasoli2021}). This method counts galaxies in spherical shells around void centers,
averaging the result over a stack of voids after rescaling all distances by $R_v$. For the SDSS sample, we apply the same estimator and radial rescaling to ensure a fully consistent comparison between observations and simulations.

Based on the Poisson estimator, we calculate the stacked density contrast profile, $\delta(r)$, for our void catalogs. The procedure involves counting galaxies within spherical shells. For a stack of $N_v$ voids, the average density in the $j$-th radial bin at a scaled distance $r_j / R_{v,i}$ from the void center is defined by

\begin{equation}
\bar{\rho^{j}} = \frac{(\sum_{i=1}^{N_v } N^{j}_{i})+1}{\sum_{i=1}^{N_v } V^{j}_{i}},
\label{eq:poisson_corrected_density}
\end{equation}

where $N_i^j$ is the number of galaxies in the shell volume $V_i^j$ of the $i$-th void, and the term $+1$ ensures the contrast is zero at the mean density, correcting for the systematic bias described in \citet{nadathur2014}. For the SDSS void sample, we apply the same estimator and radial rescaling, ensuring that the observational and simulated profiles are constructed in a fully consistent manner.

Although there is little theoretical guidance for predicting the precise shape of void density profiles, recent work has successfully described them using empirical formulas (e.g.,
\citealt{ceccarelli2013, hamaus2014, nadathur2014, Voivodic2020, Tavasoli2021}). Following this approach, we adopt the standard empirical form proposed by \citet{Barreira2015}. This parametric model captures the profile's behavior across both the underdense void interior and the surrounding compensating ridge, and it has been shown to provide accurate fits to both simulated and observational void samples. Including the SDSS profile in our analysis allows us to test whether the same functional form remains valid for real galaxies at $z\!\sim\!0$.

The model for the relative density is given by the empirical profile introduced by \citet{hamaus2014}:
\begin{equation}
 \frac{\rho_v(r)}{\bar{\rho}} = 1 + \delta_c \frac{1 - (r/r_{s1})^\alpha}{1 + (r/r_{s2})^\beta},
\label{eq:void_density_profile}
\end{equation}
where $\delta_c$ is the central density contrast, $r_{s1}$ is the zero-crossing scale, and $r_{s2}$ sets the scale at which the profile returns to zero. The exponents $\alpha$ and $\beta$ govern the inner and outer slopes of the profile, respectively.

\begin{figure*}[ht!]
    \centering
    \includegraphics[width=\textwidth]{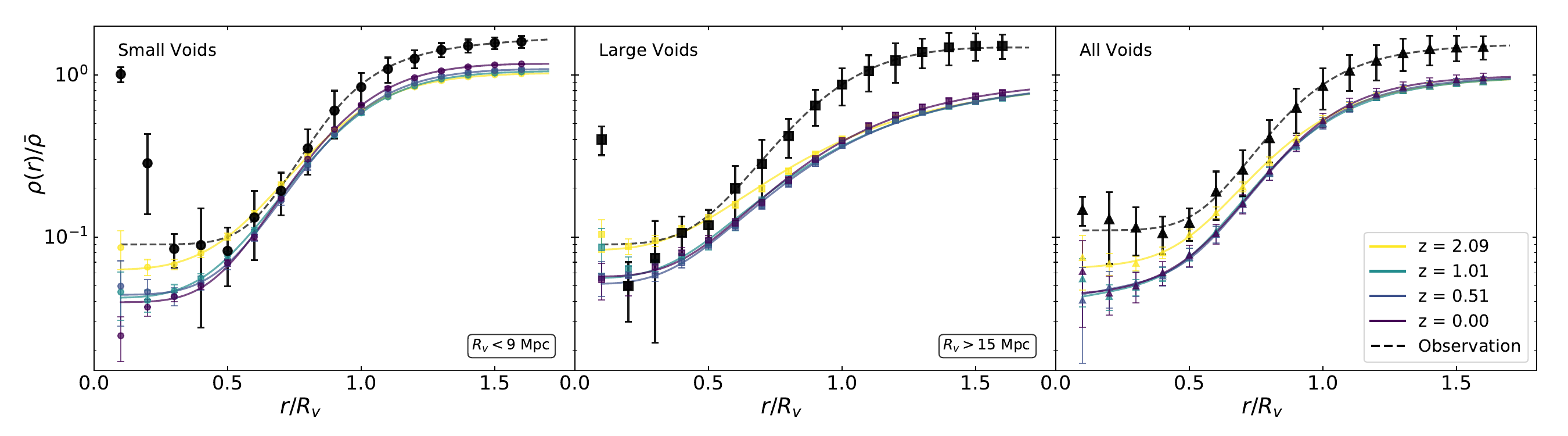} 
    \caption{Stacked density profiles of voids from the Simulation for Small, Large, and All voids at four redshifts ($z = 0.0$, $0.5$, $1.0$, and $2.0$). Markers indicate the mean values, and error bars represent the standard error of the mean. Solid curves show the best-fit solutions to Equation~(\ref{eq:void_density_profile}). For comparison, the corresponding SDSS measurements at $z \sim 0$ are overplotted as black dashed lines.}

    \label{fig:profile}
\end{figure*}

We apply this empirical model (Equation~\ref{eq:void_density_profile}) to our three void samples (Small, Large, and All voids) across the four simulation snapshots. Figure~\ref{fig:profile} shows the resulting stacked radial density profiles for each sample, with error bars representing the standard error of the mean. The SDSS profile at $z\!\sim\!0$ is overplotted as a black dashed line, using the same radial scaling and Poisson estimator, enabling a fully consistent and direct comparison between simulated and observed void environments. By fitting the model to these profiles, we obtain the
best-fit parameters ($\delta_c$, $r_{s1}$, $r_{s2}$, $\alpha$, $\beta$) for every void category and redshift; these values are compiled in Table~\ref{tab:prof_fits}, where the SDSS best-fit parameters are listed alongside the simulation results for completeness. The inclusion of the SDSS constraints provides an important low-redshift anchor for assessing the normalization of the simulated evolutionary trends.

Figure~\ref{fig:profile} reveals that the stacked density profiles at all redshifts and for all void sizes approximate an inverted top-hat shape. While void interiors remain strongly underdense ($\delta \lesssim -0.9$) in every simulated sample, the profiles evolve systematically with time: by $z = 0.0$ the cores become deeper and the surrounding ridges more pronounced, in agreement with earlier studies (e.g., \citealt{hamaus2014, cautun2014, schuster2025}). The central underdensity deepens with time—from $\delta \sim -0.91$ at $z = 2.09$ to $\delta \sim -0.98$ at $z = 0.0$ for Small voids, from $\sim -0.90$ to $\sim -0.94$ for Large voids, and from $\sim -0.92$ to $\sim -0.95$ for the All-voids sample. Small-void profiles rise steeply beyond $r/R_{\rm eff} \gtrsim 0.5$ and eventually reach positive overdensities, reflecting their embedding within overdense surroundings. In contrast,
Large-void profiles increase more gradually and approach the mean background density beyond the ridge.

This behavior is consistent with the void--in--void and void-in-cloud paradigms introduced by
\citet{Sheth2004}. Small “void-in-cloud’’ systems form within relatively high-density regions and are typically bounded by overdense ridges, reflecting their origin in the collapse of larger surrounding structures. Large “void--in--void’’ systems arise from the expansion and merging of underdense regions and therefore reside in environments that connect smoothly to the global background density. A direct quantitative comparison with previous measurements of void density profiles is not straightforward, as existing studies differ substantially in their void-finding algorithms, underlying cosmological simulations, adopted redshift ranges, and the selection criteria applied to void galaxies (e.g., \citealt{Habouzit2020, Schuster2023, Curtis2025Density}). Consequently, the numerical values
reported here should be interpreted within the context of our specific methodology rather than compared one-to-one with results in the literature.

The SDSS profile at $z\!\sim\!0$ exhibits the same qualitative structure but is systematically shifted upward relative to the simulated profiles across all void categories. At fixed scaled radius, the observed density contrast is less negative than in the simulations, indicating that real void environments are, on average, less empty. This offset is consistent with expectations arising from differences in background densities, survey selection functions, and the discrete sampling of galaxies in magnitude-limited surveys.

A particularly noteworthy feature of the SDSS profiles is the central density level.
 In all three void categories, the observed profiles show a non-zero central density,
 implying the presence of galaxies near the void centers. Such central populations are
 not reproduced in the simulations, where the profiles approach $\delta \rightarrow -1$ at 
 small radii, as expected in standard $\Lambda$CDM predictions for deep void interiors.

Several observational effects that could in principle populate void centers are mitigated in
 our analysis: galaxy redshifts are corrected to the CMB frame, which suppresses large-scale 
 redshift-space distortions, and voids intersecting survey boundaries are excluded to avoid geometric
 incompleteness. However, an important methodological limitation remains: voids in the Millennium simulation
 are identified in real space, whereas SDSS voids are defined in redshift space. This mismatch may contribute
 to part of the central density offset, as residual redshift-space distortions can shift galaxies toward void
 centers even after CMB frame corrections. A dedicated mock analysis--constructing a redshift space version of
 the simulation and reidentifying voids under SDSS like conditions--will be required to quantify this effect.

Nevertheless, the persistence of the central excess across all void size bins suggests that it is unlikely to
 arise solely from survey artifacts. This discrepancy, previously highlighted by \citet{Tavasoli2021}, indicates
 that real voids may host a small but non-negligible population of centrally located galaxies, posing an interesting
 challenge to the predictions of the $\Lambda$CDM model for galaxy formation in underdense environments.

Overall, the comparison between simulated and observed profiles demonstrates that while the global shape of
 the void density profile is broadly consistent between the two datasets, systematic offsets in normalization
 and central density persist. These differences provide valuable constraints on both the low redshift galaxy
 distribution and the physical processes governing galaxy formation within the most underdense regions of the Universe.


\begin{table}[htbp]
\centering
\caption{Best-fit parameters of the void density profile model (Equation~\ref{eq:void_density_profile}) for Small, Large, and All voids across four simulation snapshots, together with the corresponding SDSS
measurements at $z\!\sim\!0$. Simulation and observational results are presented in a consistent
format to facilitate direct comparison.}
\vspace{0.2em}
\label{tab:prof_fits}
\small
\begin{tabular*}{\linewidth}{@{\extracolsep{\fill}} l l c c c c c}
\toprule

Redshift  & Void Sample & $\delta_c$ & $r_{s1}$ & $r_{s2}$ & $\alpha$ & $\beta$ \\
\midrule

\multicolumn{7}{c}{Millennium simulation} \\
\multirow{3}{*}{0.0}  & Small & -0.96 & 1.23 & 1.07 & 4.39 & 6.18 \\
                      & Large & -0.94 & 2.01 & 1.17 & 10.0 & 3.79 \\
                      & All   & -0.95 & 2.48 & 1.04 & 2.39 & 6.15 \\
\addlinespace
\multirow{3}{*}{0.51} & Small & -0.95 & 1.33 & 1.07 & 4.12 & 6.27 \\
                      & Large & -0.94 & 2.11 & 1.23 & 10.3 & 3.52 \\
                      & All   & -0.96 & 2.05 & 1.02 & 1.10 & 5.68 \\
\addlinespace
\multirow{3}{*}{1.01} & Small & -0.95 & 1.40 & 1.08 & 3.52 & 6.18 \\
                      & Large & -0.94 & 2.02 & 1.23 & 10.0 & 3.43 \\
                      & All   & -0.96 & 2.13 & 1.03 & 1.12 & 5.30 \\
\addlinespace
\multirow{3}{*}{2.09} & Small & -0.93 & 1.50 & 1.05 & 3.17 & 6.09 \\
                      & Large & -0.91 & 2.05 & 1.23 & 10.1 & 3.25 \\
                      & All   & -0.93 & 1.45 & 1.01 & 1.41 & 5.01 \\
\midrule

\multicolumn{7}{c}{SDSS observation} \\
\addlinespace
\multirow{3}{*}{0.0}  & Small & -0.91 & 1.05 & 0.97 & 7.79 & 7.20 \\
                      & Large & -0.91 & 1.06 & 1.10 & 4.31 & 5.65 \\
                      & All   & -0.89 & 1.06 & 0.99 & 6.05 & 6.11 \\
\bottomrule

\multicolumn{7}{@{}l}{
  \parbox{\columnwidth}{
  \vspace{0.7em}
   \small \textbf{Note.} Uncertainties are estimated from the standard error of the mean of the stacked density profiles. Because the resulting errors are small and nearly constant across all void samples and redshifts, we report typical (mean) uncertainties rather than listing them for each entry. For the simulation results, the representative uncertainties are approximately $\pm 0.01$ for $\delta_c$, $\pm 0.5$ for $r_{s1}$, $\pm 0.02$ for $r_{s2}$, $\pm 0.5$ for $\alpha$, and $\pm 1$ for $\beta$. For the SDSS measurements, the corresponding uncertainties are slightly larger due to survey selection effects and sparse sampling, with typical values of $\pm 0.03$ for $\delta_c$, $\pm 0.05$ for $r_{s1}$, $\pm 0.2$ for $r_{s2}$, $\pm 3.1$ for $\alpha$, and $\pm 2.1$ for $\beta$.}
}
\end{tabular*}
\end{table}


\subsection{Dynamic Evolution of Void Galaxies}

To gain deeper insight into the dynamical evolution of cosmic voids, we examine how the spatial
distribution of void galaxies depends on both void size and redshift. For this purpose, we
characterize the locations of galaxies within their host voids using two dimensionless parameters--
the \textit{center--distance} and the \textit{mean--distance}--following the definitions introduced by
\citet{Tavasoli2021}. These two parameters capture complementary aspects of the internal void
structure--namely the radial distribution of galaxies and their local clustering properties--making
them well suited for tracing the evolution of galaxy populations within voids.

Together, these parameters allow us to quantify how galaxies populate void interiors and how this
distribution evolves as voids grow in size across cosmic time, as expected in the $\Lambda$CDM
framework. For each void, the \textit{center--distance} parameter is computed as the mean distance of
all member galaxies from the void center. Because our catalog contains voids spanning a wide range
of sizes, we normalize this parameter by the effective radius of each void, $R_{\mathrm{eff}}$, enabling a
consistent comparison across the full sample and allowing a direct connection to be made with
observational measurements at $z\!\sim\!0$.

The \textit{mean--distance} parameter provides complementary information by quantifying the degree of clustering among galaxies inside voids. To compute this quantity, we employ a Minimum Spanning Tree (MST) algorithm based on graph theory, in which all points in a set are connected by the shortest possible network without closed loops (see, e.g., \citealt{prim1957}). The MST has been widely used in astrophysical applications, including the identification of galaxy clusters and filaments and the characterization of large-scale structure (e.g., \citealt{barrow1985, bhavsar1988, plionis1992}). For each void, we construct the MST of its galaxy members and define the mean--distance parameter as the total MST branch length divided by the number of branches. As with the center--distance parameter, we normalize this value by $R_{\rm eff}$ to ensure comparability across voids of different sizes. Smaller normalized mean--distance values indicate more strongly
clustered galaxy distributions.

\citet{Tavasoli2021} showed that, at $z=0$, larger voids tend to exhibit larger center--distance values and smaller mean--distance values in both observational and simulated samples. In the present work, we extend this analysis for the first time by exploring how these trends evolve with redshift across four simulation snapshots and by comparing the $z=0$ predictions directly with SDSS measurements.
This combined simulation--observation approach enables us to assess whether the spatial
distribution of void galaxies in the simulations converges toward the patterns observed in the local Universe, and to identify any systematic offsets that may shed light on galaxy formation processes in the most underdense environments.

\begin{figure}[ht!]
    \centering
    \includegraphics[width=\columnwidth]{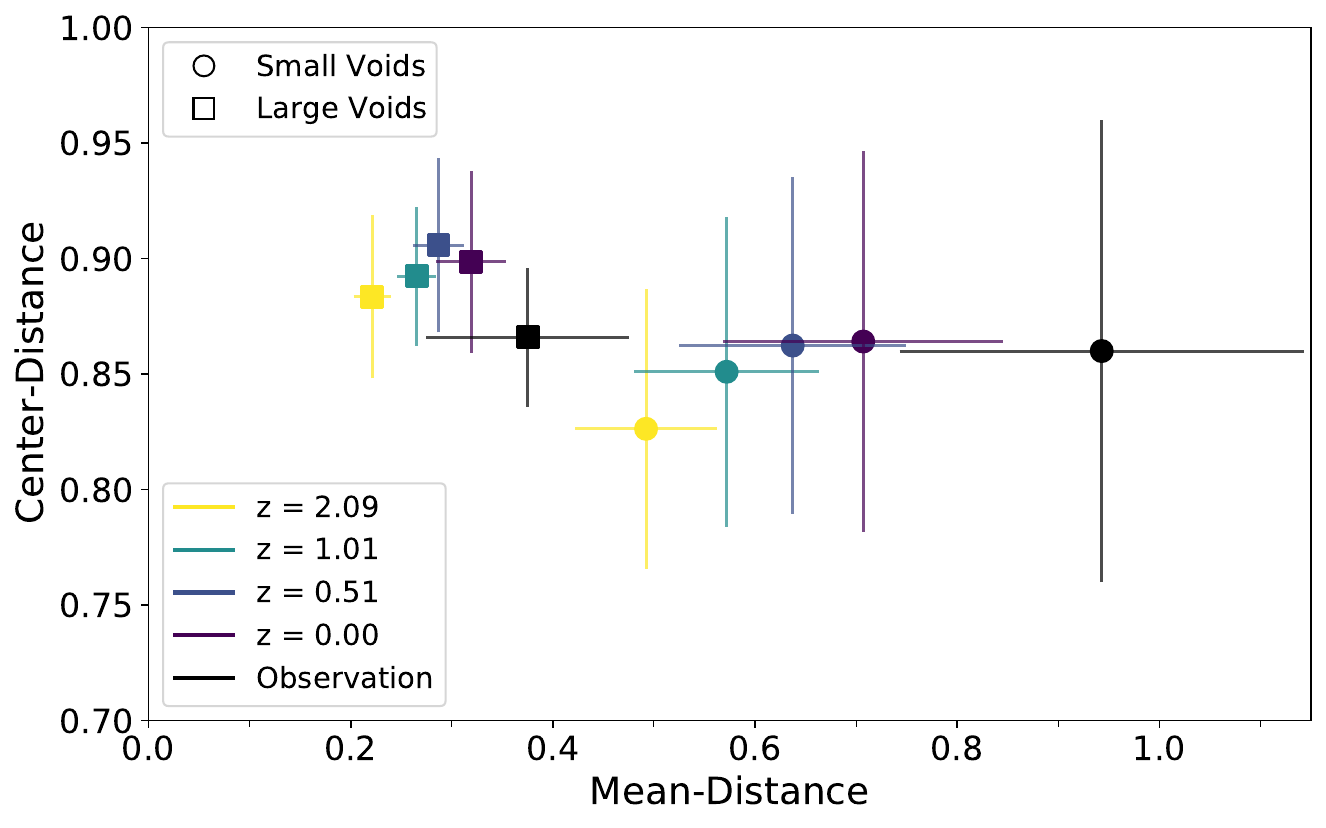}
    \caption{Mean values of the center--distance and mean--distance parameters for galaxies residing in small and large voids. Symbols show the results from four simulation snapshots, while the $z\!\sim\!0$ observational measurements are overplotted with black symbols for comparison. Error bars represent the $1\sigma$ standard deviation of each distribution.}

    \label{fig:distance}
\end{figure}

To investigate how the distribution of galaxies within voids evolves, we analyze the Small and Large void populations across four simulation snapshots. This enables us to trace how the average values of the center--distance and mean--distance parameters, together with their $1\sigma$ dispersions, change with redshift and void size within the $\Lambda$CDM framework. In addition to the simulation-based evolutionary trends, we also compare the $z=0$ predictions directly with the SDSS measurements, providing an observational benchmark for assessing whether the simulated spatial distribution of void galaxies converges toward the patterns observed in the local Universe.

Figure~\ref{fig:distance} presents the mean values of the center--distance and mean--distance parameters for both void categories. The error bars show that galaxies residing in small voids exhibit a substantially broader distribution than those in large voids, indicating that small--void environments host a more heterogeneous galaxy population. The figure also reveals a clear and persistent segregation between the two void classes: galaxies in large voids are more strongly clustered—reflected in their smaller mean--distance values—while galaxies in small voids tend to lie closer to the void center. Conversely, galaxies in large voids are typically located farther from the
center, consistent with the more diffuse and extended structure of these systems.

These qualitative trends remain stable across all redshifts examined and are also present in the $z\!\sim\!0$ observational sample, demonstrating that the spatial organization of void galaxies follows a coherent pattern that is largely independent of cosmic epoch.

Quantitatively, the simulation results show that the mean--distance parameter in large voids increases from approximately 0.20 at $z = 2.09$ to about 0.32 at $z = 0$. For small voids, this parameter evolves from roughly 0.50 to 0.70 over the same redshift interval. The center--distance parameter exhibits a similar evolutionary trend: its mean value increases from $\sim 0.88$ to $\sim 0.90$ in large voids and from $\sim 0.82$ to $\sim 0.87$ in small voids. These simulation-based trends indicate that void galaxies in both categories gradually migrate toward the void boundaries over cosmic time.

This outward migration is consistent with previous observational and simulation-based studies that found void galaxies preferentially located near the void edge \citep{Hoyle2012, Ricciardelli2014}. The observed evolution supports the picture in which the internal substructure of voids becomes less pronounced as matter flows outward, in agreement with the hierarchical void evolution scenario proposed by \citet{Sheth2004} and \citet{goldberg2004}.

At $z=0$, the observational measurements exhibit the same qualitative separation between small and large voids as seen in the simulations, but with two notable systematic offsets. First, galaxies in observed large voids display larger mean--distance values than their simulated counterparts, indicating a less strongly clustered galaxy distribution than predicted by the $\Lambda$CDM model. Second, the center--distance values show that galaxies in observational large voids tend to lie closer to the void center compared to those in the simulations. A similar behavior is present for small voids, although the difference between the observed and simulated center--distance values is considerably smaller. These trends suggest that, while the overall spatial organization of void
galaxies is consistent between observations and simulations, the observed voids--particularly the large ones--exhibit a somewhat more centrally concentrated and less clustered galaxy population.

We note that the MST mean--distance statistic is intrinsically sensitive to the number of galaxies within each void. In the present context, however, this $N$-dependence reflects a genuine physical effect rather than a sampling limitation: observational studies, including the early arguments of \citet{Peebles2001}, have shown that galaxy abundances in low-density environments are systematically lower than predicted by $\Lambda$CDM. The larger MST separations in SDSS voids therefore naturally follow from this real deficit of galaxies relative to the simulation.

Taken together, these results demonstrate that the spatial distribution of void galaxies follows a robust evolutionary pattern across cosmic time, with simulations and observations showing broadly consistent behavior but with measurable offsets that may reflect differences in galaxy formation efficiency or environmental processing within the most underdense regions of the Universe.

\begin{figure*}[!htbp]
    \centering
    \includegraphics[width=\textwidth]{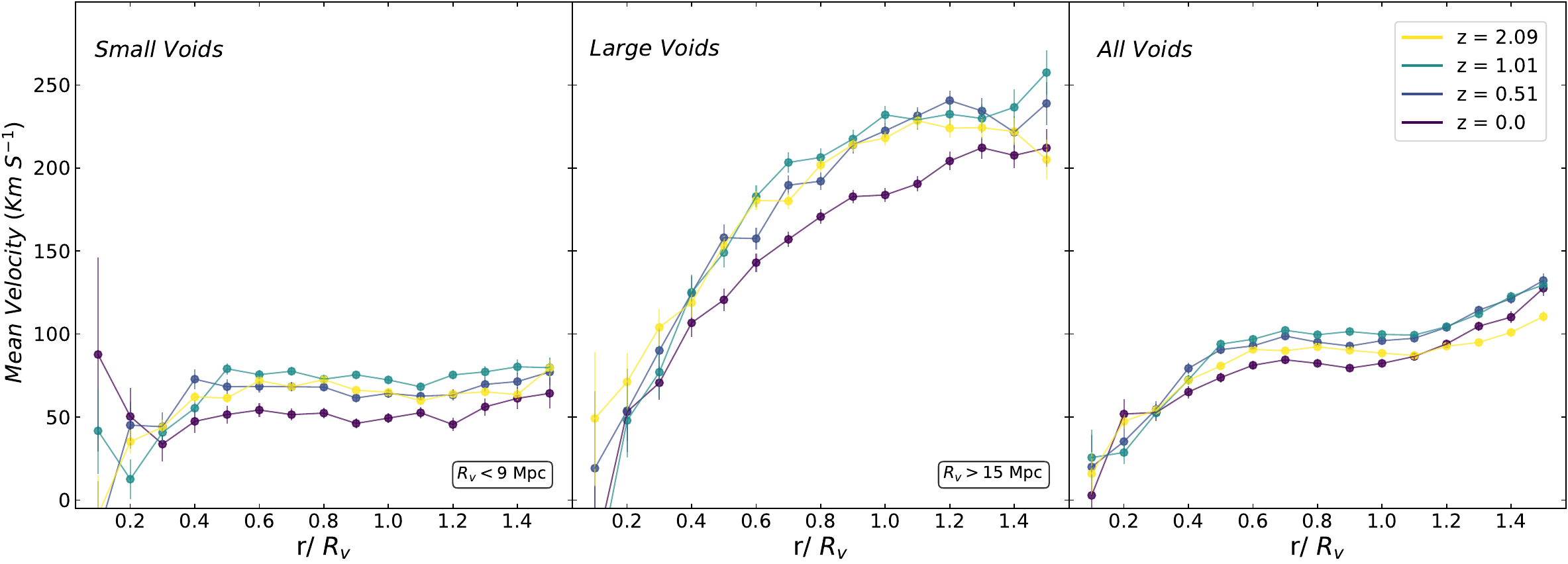} 
    \caption{Stacked radial velocity profiles for small voids, large voids, and the full void sample across four redshift intervals, derived from the Millennium simulation.} 
    
    \label{fig:velocity}
\end{figure*}

\subsection{Radial Velocity Profiles of Void Galaxies}

The peculiar velocities of galaxies within cosmic voids provide a sensitive probe of the dynamical evolution of underdense regions and the growth of large--scale structure. In the standard cosmological framework, voids expand faster than the background Universe, generating coherent outflows of matter from their interiors toward their surrounding walls. The amplitude and radial dependence of these outflows encode information about the void density profile, the surrounding environment, and the underlying cosmological model (e.g., \citealt{Sheth2004,hamaus2014,nadathur2016}).

To quantify the motion of galaxies relative to their host voids, we compute the radial component of the peculiar velocity for each galaxy. For a galaxy located at position $\vec{x}_{\rm g}$ and a void center at $\vec{x}_{\rm v}$, the displacement vector is
\begin{equation}
    \vec{r} = \vec{x}_{\rm g} - \vec{x}_{\rm v},
\end{equation}
with magnitude $r = |\vec{r}|$. The radial velocity component is then defined as
\begin{equation}
    v_r = \frac{\vec{v}\cdot\vec{r}}{|\vec{r}|},
\end{equation}
where $\vec{v}$ is the peculiar velocity of the galaxy. 

By construction, $v_r>0$ indicates that the galaxy is moving away from the void center (outflow), while $v_r<0$ corresponds to motion toward the center (infall). The sign of $v_r$ is therefore determined entirely by the dot product $\vec{v}\cdot\vec{r}$, while the denominator simply normalizes the velocity along the radial direction.

To enable stacking across voids of different sizes, we normalize the radial coordinate by the effective void radius,
\begin{equation}
    x \equiv \frac{r}{R_{\rm void}},
\end{equation}
and bin galaxies in shells of width $\Delta x = 0.1$ from the void center out to $x=1.5$. For each bin, we compute the mean radial velocity,
\begin{equation}
    \langle v_r(x) \rangle = \frac{1}{N(x)} \sum_{j\in x} v_{r,j},
\end{equation}
where $N(x)$ is the number of galaxies in the shell. This procedure is applied separately to small voids, large voids, and the combined sample, and repeated at all redshifts in our simulation.

Figure~\ref{fig:velocity} presents the resulting stacked radial velocity profiles. For small voids, the mean radial velocity remains approximately constant and positive across the entire void interior, indicating that galaxies move outward with nearly uniform speed from the center to the void boundary. The amplitude of this outflow increases systematically with redshift: the mean velocity rises from roughly $50~{\rm km\,s^{-1}}$ at $z=0$ to nearly $80~{\rm km\,s^{-1}}$ at $z=2.09$, with galaxies exhibiting larger outward motions at all radii. This behavior reflects the stronger dynamical evolution of voids at earlier cosmic times, when the Universe is more matter--dominated and the suppression of peculiar velocities by dark energy is weaker.

In contrast, large voids display a markedly different behavior. Their radial velocity profiles rise steadily with distance from the center, reaching their maximum near the void boundary. This monotonic increase is characteristic of ``void--in--void'' evolution, in which the entire region participates in coherent expansion. As with small voids, the amplitude of the outflow grows with redshift, consistent with expectations from linear theory and spherical expansion models.

The combined sample exhibits an intermediate behavior: the mean radial velocity increases slowly at small radii, flattens near the void boundary, and then rises again beyond $x\sim 1$. This mixed trend reflects the superposition of small and large voids in the stacked profile, demonstrating the importance of size--segregated analyses when interpreting void dynamics.

Our radial velocity profiles show a clear size--dependent dichotomy that aligns broadly with the picture established by \citet{hamaus2014}. For large voids ($R_v>15$ Mpc), we recover the characteristic monotonic outflow rising to a maximum near the void boundary, consistent with undercompensated ``void‑-in‑-void'' expansion. The amplitude and shape of these profiles are in good agreement with their results, indicating that large-‑void kinematics are robust across different simulations, tracer populations, and void‑-finding methods.

For small voids, however, our results differ: we find a nearly constant positive outflow throughout the interior, whereas \citet{hamaus2014} report a sign change (infall) near the boundary for the smallest voids, indicative of overcompensation (``void--in--cloud''). This discrepancy likely stems from differences in tracer selection (galaxies vs. dark matter particles), the exclusion of voids with $R_v<7$ Mpc from our catalog, and the use of distinct void finding algorithms (\citealt{Aikio1998} vs. ZOBOV). Nevertheless, both studies reinforce the key role of void size in governing internal kinematics.

\begin{figure}[ht!]
    \centering
    \includegraphics[width=\columnwidth]{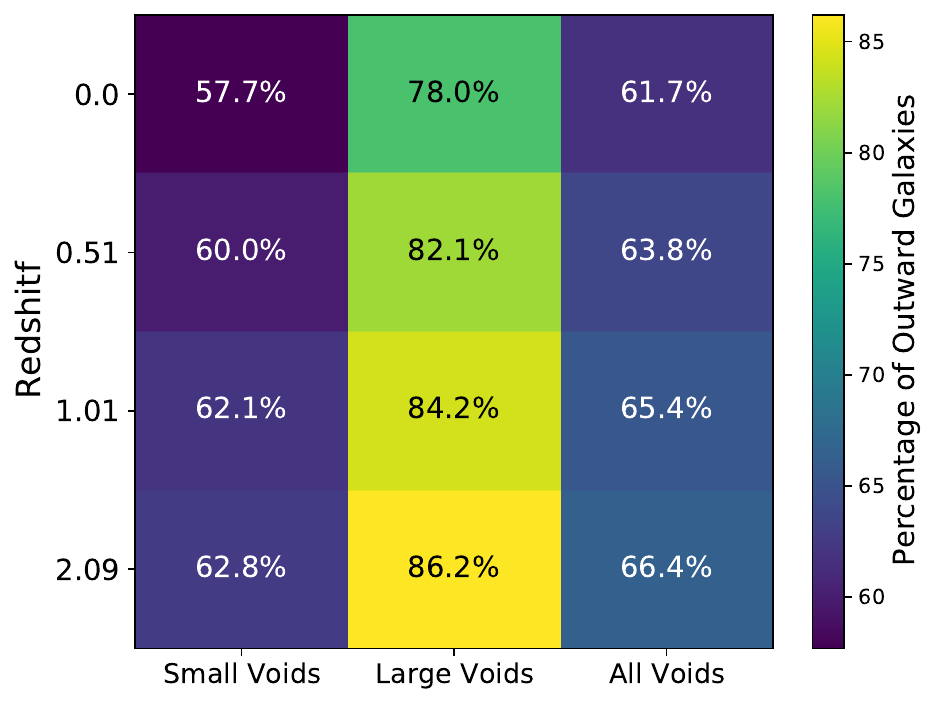}
   \caption{Heat--map showing the fraction of outward--moving galaxies for small voids, large voids, 
and the full void sample across four redshift intervals.}
    \label{fig:heat}
\end{figure}

To further quantify the directional coherence of galaxy motions within voids, we compute the fraction of outward--moving galaxies, defined as the percentage of objects with $v_r>0$ within each void category. Figure~\ref{fig:heat} presents these results in the form of a heat--map, illustrating the redshift evolution of the outward--motion fraction for small voids, large voids, and the full stacked sample. For small voids, this fraction increases modestly from $57.7\%$ at $z=0$ to $62.8\%$ at $z=2.09$, indicating a mild enhancement in coherent outflow at earlier cosmic times. Large voids exhibit a substantially stronger trend, with the outward fraction rising from $78.0\%$ to $86.2\%$ over the same redshift interval, reflecting the more vigorous and self--similar expansion characteristic of mature ``void--in--void'' systems. The full void population shows an intermediate behavior, increasing from $61.7\%$ to $66.4\%$ between $z=0$ and $z=2.09$.

These population--level statistics reinforce the conclusions drawn from the radial velocity profiles. The significantly higher outward fractions in large voids, together with their stronger redshift evolution, demonstrate that galaxy migration away from the void center is more efficient in these environments. In contrast, the weaker evolution observed in small voids is consistent with the ``void--in--cloud'' scenario, where the surrounding overdense regions partially suppress coherent expansion. Taken together, the radial velocity profiles and the outward--motion fractions highlight the sensitivity of void galaxy kinematics to both void size and cosmic epoch, and align closely with theoretical expectations and previous simulation--based studies (e.g., \citealt{Sheth2004,hamaus2014,cautun2018}).

We note that the radial velocity profiles and outward--motion fractions presented in this section can
only be computed directly in the simulation, where the full three--dimensional peculiar velocity
field of each galaxy is available. In observational data, only the line--of--sight component of the
velocity can be inferred from redshifts, and the transverse components required to evaluate
$v_r = (\vec{v}\cdot\vec{r})/|\vec{r}|$ are not accessible. As a result, galaxy--by--galaxy radial
velocities and the corresponding outward--motion statistics cannot be measured for the observational
void sample. Instead, dynamical information in real surveys must be extracted indirectly from
redshift--space distortions in the void--galaxy distribution, which encode the statistical imprint of
the same velocity fields analyzed here.

Overall, these results provide a coherent dynamical picture in which void galaxies participate in
systematic outward flows whose amplitude and radial structure depend sensitively on void size and
cosmic epoch.

Despite these observational limitations, the simulation-based velocity profiles offer essential
physical insight into the dynamical processes that shape the void--galaxy distribution observed in
redshift space.

\section{Conclusion}

In this work, we examined the structural, photometric, and dynamical evolution of cosmic voids and their galaxy populations, with particular emphasis on the role of void size. Using void catalogs constructed from four Millennium Simulation snapshots ($z=0$--$2.09$) and an observational counterpart from the SDSS at $z<0.04$, we performed a unified analysis of void demographics, galaxy properties, and internal dynamics across cosmic time.

The comparison between simulated and observed voids reveals a consistent evolutionary picture. High--redshift voids contain more galaxies, exhibit higher number densities, and lack the largest systems seen at $z\sim 0$, in agreement with hierarchical void growth. The SDSS voids show fewer galaxies than the simulated $z=0$ snapshot, consistent with the observed ``void phenomenon,'' while the distributions of effective radii broadly agree. Trends in void radius, galaxy counts, and density contrast demonstrate that simulated voids become progressively emptier toward low redshift, with apparent offsets between the SDSS and simulation arising primarily from differences in background density normalization.

The luminosity function of void galaxies exhibits coherent evolution across $0<z<2.09$: $M^{*}$ fades and $\alpha$ becomes shallower toward low redshift. While galaxies in Large voids are systematically brighter at any given epoch, the evolutionary amplitude of both $M^{*}$ and $\alpha$ is larger for Small voids, indicating that galaxy populations in smaller underdensities undergo more pronounced changes over cosmic time. The SDSS measurements at $z\!\sim\!0$ reproduce the same size--dependent hierarchy, albeit shifted toward less negative $M^{*}$ and $\alpha$, likely reflecting survey selection and environmental differences. These trends, together with the evolution of color and sSFR, support an environmentally modulated void--downsizing scenario.

The stacked radial number--density profiles show a broadly universal shape when scaled by $R_v$, with deepening central underdensities and increasingly pronounced compensating walls toward $z=0$. Small voids exhibit sharper compensating walls, while Large voids show smoother transitions consistent with void--in--void evolution. The SDSS profiles share the same qualitative form but are shifted to higher densities and contain a non-zero central population absent in the simulations, reflecting sampling and background--density differences.

The spatial distribution of void galaxies reveals a persistent segregation between Small and Large voids: galaxies in Large voids are more strongly clustered and lie farther from the void center, whereas those in Small voids occupy more heterogeneous and centrally concentrated configurations. Both the center--distance and mean--distance parameters increase toward low redshift, indicating a gradual outward migration of galaxies as voids evolve. At $z\!\sim\!0$, the SDSS measurements reproduce these qualitative trends but show a less clustered and more centrally concentrated galaxy distribution than the simulations, particularly in Large voids. Notably, the SDSS profiles exhibit a non-zero central galaxy population---absent in the $\Lambda$CDM predictions---which represents a potential challenge for standard models of void evolution and galaxy formation in extremely underdense environments.

The dynamical analysis shows coherent outward flows in all void categories, with amplitudes that are larger at higher redshift. Small voids exhibit nearly flat, positive radial--velocity profiles, while Large voids show steadily rising outflows characteristic of void--in--void evolution. The fraction of outward--moving galaxies follows the same hierarchy. Because full three--dimensional velocities are unavailable observationally, these results are simulation--only, but they provide the physical interpretation for the redshift--space distortions observed in real surveys.

We note that the observational comparison is based on a single low‑-redshift SDSS slice containing a limited number of voids (\(N_{\rm void}=93\); Table~\ref{tab:voidcatalog}). Although standard resampling techniques such as jackknife or bootstrap are commonly used to estimate observational uncertainties, the small sample size and the spatial clustering of voids in the SDSS footprint violate the independence assumptions of these methods and can yield misleading error estimates; accordingly, we do not apply jackknife or bootstrap resampling to the observational catalog in this study. A full characterization of cosmic variance and selection effects would require multiple independent simulation volumes or dedicated SDSS‑-like mock catalogues and/or substantially deeper and wider observational data, which are beyond the scope of the present work. Where appropriate (e.g., for outward‑-motion fractions) we report simple binomial confidence intervals as a transparent, well‑defined statistical descriptor.

Together, our findings present a coherent framework in which void size, cosmic epoch, and environment jointly regulate the structural, photometric, and dynamical evolution of voids and their galaxies. Future comparisons with a broader suite of simulations and deeper high--redshift surveys will be essential for assessing the generality of the evolutionary trends reported here. In particular, extending this analysis to full hydrodynamic simulations such as IllustrisTNG \citep{nelson2019} and EAGLE \citep{schaye2015}, as well as simulations incorporating updated cosmological parameters from Planck \citep{ade2016}, will provide a more stringent test of the physical origin and universality of these results--especially regarding the unexpected central galaxy population in observed voids.

\bibliography{sample701}

\end{document}